\documentclass{elsart}

\usepackage{epsfig,amsmath,amssymb}

\def\cmds{cm$^{-2}$\,s$^{-1}$}
\def\cmts{cm$^{-3}$\,s$^{-1}$}

\begin{document}

\begin{frontmatter}

\title{Neutron background at the Canfranc Underground Laboratory
and its contribution to the IGEX-DM dark matter experiment}

\author{J.M.~Carmona}, \author{S.~Cebri\'an},
\author{E.~Garc\'{\i}a}, \author{I.G.~Irastorza},
\author{G.~Luz\'on}, \author{A.~Morales\corauthref{label}},
\author{J.~Morales}, \author{A.~Ortiz de Sol\'orzano},
\author{J.~Puimed\'on}, \author{M.L.~Sarsa}, \author{J.A.~Villar.}
\corauth[label]{Deceased.}

\address{Laboratory of Nuclear and High Energy Physics,
University of Zaragoza, 50009 Zaragoza, Spain}

\begin{abstract}
A quantitative study of the neutron environment in the Canfranc
Underground Laboratory has been performed. The analysis is based
on a complete set of simulations and, particularly, it is focused
on the IGEX-DM dark matter experiment. The simulations are compared
to the IGEX-DM low energy data obtained with different shielding
conditions. The results of the study allow us to conclude, with
respect to the IGEX-DM background, that the main neutron population,
coming from radioactivity from the surrounding rock, is
practically eliminated after the implementation of a suitable
neutron shielding. The remaining neutron background (muon-induced
neutrons in the shielding and in the rock) is substantially below
the present background level thanks to the muon veto system. In
addition, the present analysis gives us a further insight on the
effect of neutrons in other current and future experiments at the
Canfranc Underground Laboratory. The comparison of simulations
with the body of data available has allowed to set the flux of
neutrons from radioactivity of the Canfranc rock, ($3.82 \pm
0.44)\times 10^{-6}$\,\cmds, as well as the flux of muon-induced
neutrons in the rock, $(1.73 \pm 0.22(stat) \pm 0.69(syst))\times
10^{-9}$\,\cmds, or the rate of neutron production by muons in the
lead shielding, ($4.8 \pm 0.6 (stat) \pm 1.9 (syst))\times
10^{-9}$\,\cmts.
\end{abstract}

\end{frontmatter}

% main text
The search for dark matter is one of the biggest challenges of
modern cosmology. Last observational data \cite{map,sn,sdss} are
compatible with a flat accelerating Universe with a matter content
of around 30\%, where about 25\% is dark matter, mostly in the
form of non-baryonic cold dark matter. Even though the Standard
Model (SM) of particle physics does not offer a proper candidate
satisfying all the needed requirements, its supersymetric
extensions open a new world of particles. Experiments looking for
cold, non-baryonic, weakly interacting, massive neutral particles
beyond the SM
---generically called WIMPs--- supposedly forming this missing
matter of the Universe are obviously of most relevance for
particle physics and cosmology and so the number of such attempts
---with a large variety of techniques, detectors and targets--- is
quite numerous \cite{reviewmorales}.

In experiments intended for WIMP direct detection, such as IGEX
Dark Matter (IGEX-DM), galactic WIMPs could be detected by means
of the nuclear recoil they would produce when scattered off target
nuclei of suitable detectors. The sensitivity of these experiments
has been continuously increasing since the first searches, more
than fifteen years ago, thanks to higher levels of radiopurity of
detectors, discrimination techniques and a deeper understanding of
the various sources of radioactive background. The study of the
background of these experiments is then a very important activity
in their development. Even though the neutron component of the
background has been always a matter of concern for shallow site
experiments \cite{Silva}, in deep underground locations it used to
be by far below the level of the typical gamma background. More
recently, however, the extreme level of radiopurity and the
expertise achieved in low background techniques, as in the case of
IGEX \cite{igex2000,igex2001,wavelets}, has reduced the raw
background to levels where the neutron contribution can be of
great importance. These facts, together with the recent development of
discrimination techniques to disentangle nuclear recoils produced
by particle dark matter interactions from electron recoils
generated by the typical gamma or beta background
\cite{cdms,edelweiss,cresst,rosebud} have made the neutrons to
remain as the real worrisome background for WIMPs
\cite{chardin,wulandari,kudry,perera}. They can produce nuclear
recoils ($<$\,100\,keV) in the detector target nuclei which would
mimic WIMP interactions. Simple kinematics tells that in the case
of germanium, neutrons of 1(5)\,MeV could elastically scatter off
germanium nuclei producing recoils of energies up to 54(268)\,keV.

\begin{figure}[htb]
\centerline{\includegraphics[width=10cm]{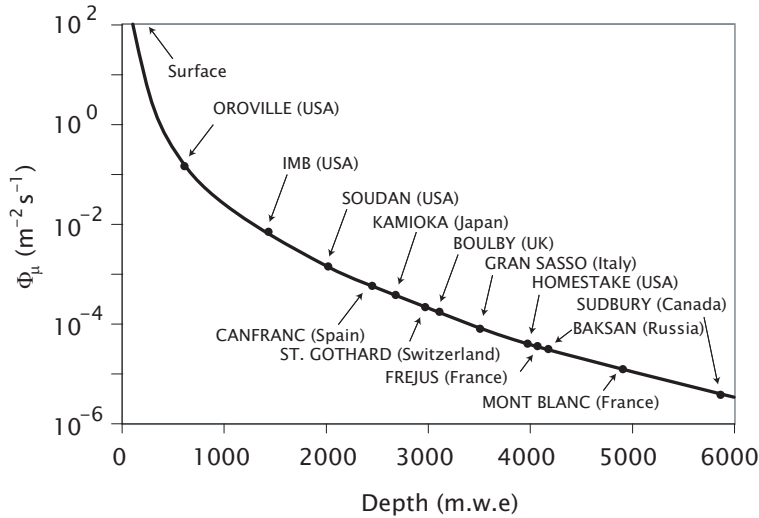}}
\caption{Dependence of muon flux with depth, showing the location
of the Canfranc Underground Laboratory with respect to other
underground facilities.} \label{canfranc} \smallskip
\end{figure}

The present work deals with the study of the neutron background in
the underground environment of the Canfranc Underground Laboratory
(LSC) (2450\,m.w.e., see Fig.~\ref{canfranc}), and in particular
of its fast component with effect on the IGEX-DM experiment. In
Section 1 the IGEX-DM experiment is briefly described, especially
in the aspects relevant to the present analysis, such as, for
instance, the lead and polyethylene shieldings and the active veto
system. The body of data available for this study is also
presented. In Section 2, a brief outline of the simulations
performed and their technical details are given. Sections 3, 4 and
5 present the results from the simulations and their comparison
with experimental data concerning the three relevant neutron
populations. The final conclusions are gathered in Section 6.

\section{The IGEX Dark Matter Experiment}

The IGEX experiment, originally optimized for detecting $^{76}$Ge
double beta decay, has been described in detail elsewhere
\cite{igexdbd}. One of the enriched (86\% in $^{76}$Ge) IGEX
detectors (named RG-II) of 2.2\,kg (2.0\,kg active mass) is being
used to look for WIMPs interacting coherently with the germanium
nuclei. Its full-width at half-maximum (FWHM) energy resolution is
2.37\,keV at the 1333\,keV line of $^{60}$Co, and the low energy
long-term energy resolution (FWHM) is 1\,keV  at the 46.5\,keV
line of $^{210}$Pb. The lines of an external $^{22}$Na source and
the excited X-rays of Pb have been used for periodic energy
calibrations at high and low energies.

We refer to papers \cite{igex2000,igex2001,wavelets} where the
latest results of the experiment regarding dark matter searches
are presented and where the aspects related to the experimental
set-up, shielding, data acquisition system, etc. are described.
The threshold of the experiment is 4\,keV and the raw background
registered in the region just above the threshold is of a few
tenths of counts/(kg\,day\,keV), the lowest raw background (i.e.\
with no nuclear recoil discrimination mechanism) ever achieved in
this energy range.

\begin{figure}[htb]
\centerline{\includegraphics*[width=10cm]{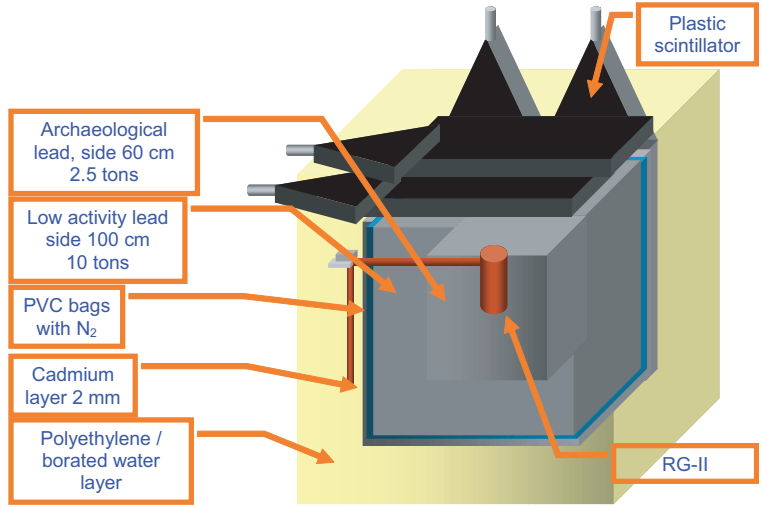}}
\vspace{0.2cm} \caption{Sketch of the experimental set-up. Four
more vetoes, covering two lateral sides of the cube, have been
removed for better visualization of the figure.} \label{drawing}
\smallskip
\end{figure}

Particularly interesting for the purpose of the present work is
the shielding used in the experiment. Leaving more detailed
information for the references previously mentioned, it can be
described as follows. The innermost part consists of about 2.5
tons of 2000-year-old archaeological lead (having $<$\,9\,mBq/kg
of $^{210}$Pb($^{210}$Bi), $<$\,0.2\,mBq/kg of $^{238}$U, and
$<$\,0.3\,mBq/kg of $^{232}$Th) forming a cubic block of 60\,cm
side. The germanium detector is fitted into a precision-machined
chamber made in this central core, which minimizes the empty space
around the detector available to radon. Nitrogen gas, at a rate of
140\,l/hour, evaporating from liquid nitrogen, is forced into the
small space left in the detector chamber to create a positive
pressure and further minimize radon intrusion. The archaeological
lead block is surrounded by 20\,cm of lead bricks made from
70-year-old low-activity lead ($\sim$\,10\,tons) having
$\sim$\,30\,Bq/kg of $^{210}$Pb. The whole lead shielding forms a
1\,m side cube, the detector being surrounded by not less than
40-45\,cm of lead (25\,cm of which is archaeological).  Two layers
of plastic seal this central assembly against radon intrusion and
a 2-mm-thick cadmium sheet surrounds the ensemble. Specially
relevant for the present work are the cosmic muon vetoes
(BC408 plastic scintillators)
covering the top and three sides of the shield. Its effect on
the background reduction in the low energy region will be stressed
in Section 3, where the neutrons induced by muons in the shielding
are studied. Finally, an external neutron moderator (made of
polyethylene bricks and borated water tanks) surrounds the whole
set-up. Its thickness has been changed in several occasions,
giving rise to independent sets of data whose differences will be
particularly useful in this analysis. See a sketch of
the set-up in Fig.~\ref{drawing}.

\begin{figure}[htb]
\centerline{\includegraphics[width=10cm,angle=0]{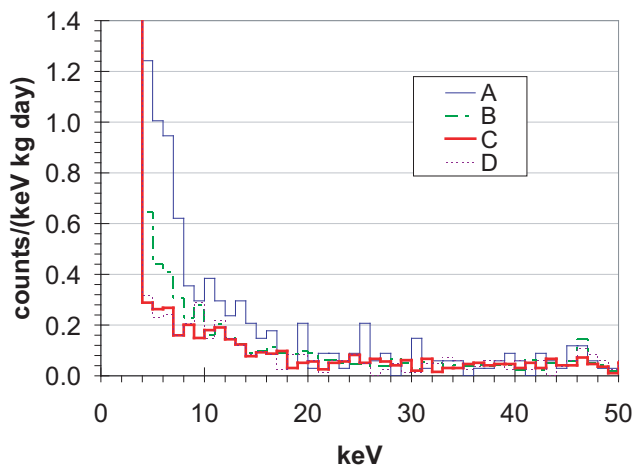}}
\vspace{-0.5cm} \caption{Low energy region of the measured spectra
for each set of data, having a different thickness of neutron
moderator: A, 0\,cm; B, 20\,cm; C, 40\,cm and D, 80\,cm.}
\label{spectrasets} \bigskip
\end{figure}

The body of data to be used in the present work has been divided
in four different sets (A, B, C and D) according basically to the
thickness of the outer neutron moderator wall used in the
shielding (0, 20, 40 and 80\,cm respectively). Some parameters
concerning each set of data are quoted in Table \ref{sets}, and
the low energy spectral shapes are plotted in
Fig.~\ref{spectrasets}. The sets B and C correspond mostly to the
data already presented in Refs.~\cite{igex2000} and
\cite{igex2001} respectively with added statistics. The data sets
A and D were taken more recently to complete the present analysis.
A word of caution must be said concerning the oldest data set B
since there are more differences in the experimental set-up with
respect to the other sets than just the neutron shielding:  four
Ge detectors were in the same lead shielding and the outer
location of the corresponding four liquid N$_{2}$ dewars left two
of the shielding sides without moderator. Therefore, set B will be
included in some plots just to have another intermediate situation
between the two extreme cases A and D, but no quantitative conclusion about
neutrons will be drawn from it. In the other three data sets (A, C
and D) the experimental set-up has been exactly the same except
for the thickness of the moderator wall.

\begin{table}[htb]
\caption{Main features of the data sets A, B, C and D: statistics
time, thickness of the neutron moderator and measured average
background level (in anticoincidence with the veto system) from 4
to 10\,keV.}
\begin{tabular}{c|cccc} \hline
 & A & B & C & D   \\ \hline
thickness of moderator (cm) & 0 & 20 & 40 & 80 \\
statistics time (days) & 17 & 118 & 97 & 41 \\
background [counts/(kg\,keV\,day)] & 0.74(6) & 0.39(2) & 0.22(1) & 0.24(2) \\
\hline
\end{tabular}
\label{sets} \bigskip
\end{table}

The differences of the spectra shown in Fig.~\ref{spectrasets}
appear to be consistent with the corresponding thicknesses of the
neutron moderator wall used. The thicker the wall, the lower the
background, excepting the last step from 40 to 80\,cm which
basically does not improve it further. The successive reduction
from the first batch of IGEX-DM data \cite{igex2000} was tentatively
interpreted as due to the disappearance of neutrons after the
increase of the neutron moderator thickness \cite{igex2001}. This
interpretation is confirmed in the present work, where a general
and quantitative study of the effect of neutrons on IGEX-DM has been
performed.

\section{General outline of the simulations performed}

In deep underground locations neutrons have two possible origins:
either they are produced by natural radioactivity in the material
surrounding the detector or they are induced by muons, the only
cosmic component still present there. The relevant
muon-induced neutrons are produced in the last few meters of rock
surrounding the laboratory or in the lead shielding itself, the
difference being extremely important, because the latter can be
rejected by the muon veto system while most of the former cannot.
Therefore, three different populations of neutrons have
been simulated: muon-induced neutrons in the shielding itself,
muon-induced neutrons in the surrounding rock and neutrons from
radioactivity in the rock. The description of the simulation in
each case and the corresponding results will be presented
independently in the following sections.

In underground locations such as the LSC (2450 m.w.e.), where the
flux of cosmic muons has been notably suppressed, the neutrons
from radioactivity in the rock are usually dominant in comparison
with the muon-induced neutrons \cite{Heusser}. However, because of
the energies of the neutrons from radioactivity ($<$\,10\,MeV)
they can be more easily and effectively shielded than those
originated by muons (whose energies go up to several hundreds of
MeV). More quantitative discussions are left for the following
sections.

Neutrons produced by radioactivity in the shielding materials have
not been included in the simulations since, owing to the extreme
radiopurity of such materials, its contribution has been
estimated to stand three orders of magnitude below the present
level of background.

Two different kinds of simulations have been performed. The first
one is that of the transport and interaction of the different
populations of neutrons through the IGEX geometry. These
simulations have been performed with the GEANT4 code \cite{geant4}
using the high precision neutron data library G4NDL3.5. The
incoming neutrons have been launched isotropically and uniformly,
either from the outer surface of the shielding in the case of
neutrons from the rock or from the whole lead volume in the case
of neutrons induced in the shielding. Spectral samplings are
described and justified later. The spectra shown as a result of
these simulations correspond always to the energy depositions of
the neutrons in the active volume of the germanium detector
(nuclear recoils) corrected for a quenching factor of 0.25 in Ge,
so that they can be directly compared with the experimental data
shown always in electron-equivalent energy. On the other hand,
some complementary simulations to study the neutron production by
muons in several generic situations have been performed with the
FLUKA code \cite{fluka}, of proven reliability when hadronic
processes are involved. These simulations provide the neutron
spectrum exiting a layer of material when a known flux of muons
traverses it, as well as the total neutron yield. We note that in
our simulation results we will not consider systematic errors 
coming from the specific implementation of physical processes
by these Monte Carlo codes.

\section{Muon-induced neutrons in the shielding}

The production of neutrons by muons interacting in matter is not a
simple subject, many different processes coming into play. The
total neutron flux and spectral shape depend on
the target material as well as on the muon energy, and therefore
on the depth. Following the practical approach of
Refs.~\cite{perera,golwala} we distinguish two different sets of
muon-induced neutrons, according to their typical energies: high
energy (HE) neutrons going up to several hundreds of MeV which are
produced basically by hadron showers originated by the muons, and
 medium energy (ME) going up only to 20-30\,MeV,
produced by photonuclear reactions related to electromagnetic
showers, capture of slow muons (of very little importance at deep
places), elastic interaction of muons with neutrons inside nuclei
and secondary neutrons produced after one of the previous
processes. Even though neutrons are produced in any shielding
material we will focus on those originated in lead owing to its
higher neutron yield and to its closeness to the detector. In
order to determine their spectral shape we have simulated with the
FLUKA code the outcome of muons passing through lead. The energy
spectrum of muons corresponding to the 2.5\,km.w.e.\ depth of the
LSC has been sampled following Ref.~\cite{Lipary}.

\begin{figure}[htb]
\centerline{\includegraphics[angle=270,width=14cm]{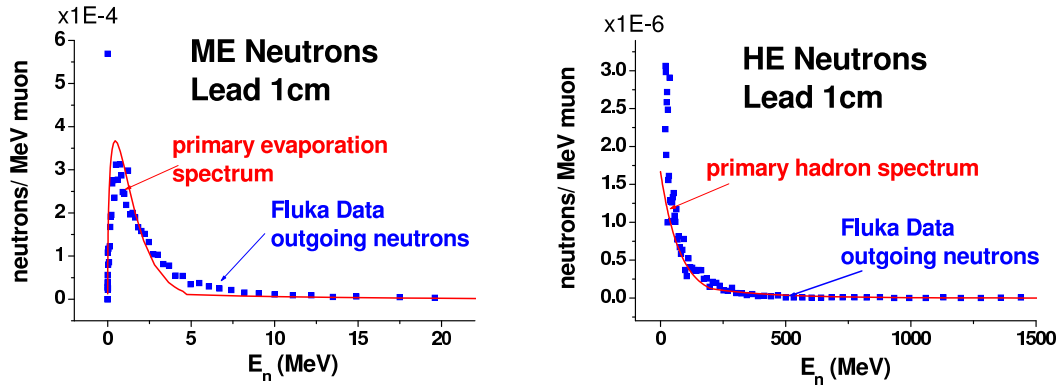}}
\caption{Energy spectrum of muon-induced neutrons in lead
according to the FLUKA simulation (squares) compared to the
analytical approximations used for ME and HE (solid lines).}
\label{flukalead} \smallskip
\end{figure}

In Fig.~\ref{flukalead} the result of such simulation for 1\,cm of
lead is shown for both ME and HE populations. Such a small
thickness avoids the production of secondary neutrons and spectral
deformations due to neutron transport through the lead slab. The
obtained spectra, which are therefore the production spectra for
primary neutrons, turns out to be in satisfactory agreement with
the analytical approximation proposed in \cite{Silva} for ME
neutrons in lead,
\begin{equation}
\frac{dN}{dE}=\left\{\begin{array}{lr}0.812 \,E^{5/11}
\exp(-E/1.22) &
      \mbox{for } E<4.5\,\mathrm{MeV} \\ 0.018 \exp(-E/9) & \mbox{for }
      E>4.5\,\mathrm{MeV} \end{array} \right.
      \label{muinme}
\end{equation}
and, in the case of HE neutrons, with the analytical formula
\begin{equation}
\frac{dN}{dE}=\left\{\begin{array}{lr}6.05 \exp(-E/77) &
      \mbox{for } E<200\,\mathrm{MeV} \\ \exp(-E/250) & \mbox{for }
      E>200\,\mathrm{MeV} \end{array} \right.
    \label{muinhe}
\end{equation}
which was originally proposed in \cite{perera,golwala} for
describing the HE neutron spectrum in rock. In these formulas $E$
is the emitted neutron energy (in MeV).

This simulation underestimates the total neutron yield since
showers, main sources of secondary neutrons, do not develop in
1\,cm of lead. The neutron yield obtained after 1\,cm of lead,
$8.8\times 10^{-5}$\,(g/cm$^{2}$)$^{-1}$ per muon, including only
primary neutrons, agrees with the numbers presented in
\cite{wulandari,kudry} for neutrons produced in inelastic processes. The
total neutron yield including also secondary neutrons can be
estimated by a second simulation with a thicker lead layer
(35\,cm). The obtained number, $1.7\times
10^{-3}$\,(g/cm$^{2}$)$^{-1}$ neutrons per muon, is pretty close
to the value estimated in \cite{chardin} for the total neutron
yield. Therefore, secondary neutrons coming from showers dominate
the muon-induced neutron population. These secondary neutrons
follow mainly the evaporative spectrum described by
Eq.~(\ref{muinme}) while primary spallation neutrons can be
sampled by Eq.~(\ref{muinhe}).

\begin{figure}[htb]
\centerline{\includegraphics[width=7cm,angle=270]{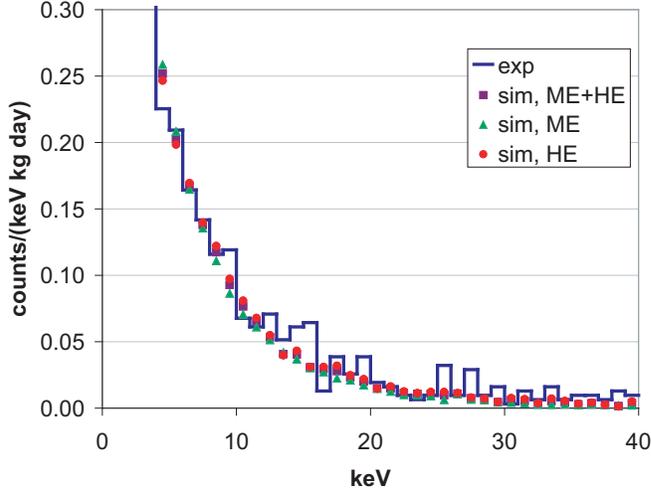}}
\vspace{-0.2cm} \caption{Comparison of the measured spectrum of
vetoed events (labelled $exp$) with the simulated spectra of
electron-equivalent energy deposited in the detector by
muon-induced neutrons in lead, assuming the HE spectrum, the ME
spectrum and a weighted combination (50\% ME, 50\% HE).}
\label{nshield} \smallskip
\end{figure}

Simulations of neutrons isotropically distributed in the full lead
volume of the IGEX geometry with spectra given by
Eqs.~(\ref{muinme}) and (\ref{muinhe}) were performed using
GEANT4. The resulting spectra at the detector turn out to be
essentially independent of the relative weight of ME neutrons and
HE neutrons, and is shown in Fig.~\ref{nshield}. The difference
between the energy distributions produced using these two spectra
(HE and ME) is small enough to make the relative abundance of HE
and ME neutrons in the input spectrum not crucial for our
purposes.

Let us compare now these neutron simulations with experimental
data. As stated before, background events coming from muon-induced
neutrons in the shielding can be rejected by the muon veto system.
The rejected events are those appearing in the experimental
coincidence spectrum between detector and vetoes. This can be
compared with the simulation results to get the normalization of
muon-induced neutrons in the IGEX shielding, assuming that no
other kind of events different from muon-induced neutrons are
present in the coincidence spectrum. We have checked that random
coincidences and events produced by direct interaction of muons in
the detector are at least 2 orders of magnitude below the total
rate of the coincidence spectrum [0.16\,counts/(keV\,kg\,day) in
4-10\,keV], so that it can be safely compared with the simulation.
This experimental spectrum is independent of the thickness of the
polyethylene wall. In Fig.~\ref{nshield}, the normalization of the
spectrum produced by the simulation has been adjusted to fit the
experimental data. From that fit the neutron production by muons
in the IGEX lead can be deduced to be $(4.8\pm 0.6(stat)\pm
1.9(syst))\times 10^{-9}$\,cm$^{-3}$\,s$^{-1}$. The systematic
error accounts only for the unknown fraction of HE and ME
neutrons and it has been estimated by calculating the neutron
production in the extremal cases of all neutrons having either 
the HE or the ME spectrum. Taking into consideration the effect 
of the evaluated rates of vetoed events which do not correspond 
to neutrons, the corresponding error is estimated to be of 0.4\%;
this systematic error is negligible with respect to
that due to the unkown fraction of HE and ME neutrons. Using the total
neutron yield previously obtained by the muon simulation together with the
estimated neutron production, the total muon flux crossing the 
shielding can be estimated and the known errors propagated, giving 
$(2.47 \pm 0.31(stat) \pm 0.98(syst)) \times 10^{-7}$\,cm$^{-2}$\,s$^{-1}$.

The obtained muon flux is of the expected order of magnitude at
2450\,m.w.e., and in fact it is in perfect agreement with previous
scintillator measurements at the LSC site, which gave a flux of
$2\times 10^{-7}$\,cm$^{-2}$\,s$^{-1}$ \cite{igex2001}, therefore
showing the reliability of our numerical simulations.

It is worth noting that the contribution of the muon-induced
neutrons generated in the shielding to the background levels is
not negligible, but thanks to the veto system it is significantly
reduced. The veto counting rate, shown in Fig.~\ref{nshield}, in
the low energy region (4-10\,keV) in IGEX-DM is 0.16 counts/(keV
kg day). For a veto efficiency $\epsilon$, such a rate would imply
a background contribution of $0.16\times (1-\epsilon)/\epsilon$
counts/(keV kg day). The efficiency of the veto system is
estimated to be well above 92$\%$  so that the contribution of
these neutrons is less than $1.3 \times 10^{-2}$ counts/(keV kg
day). This value is more than one order of magnitude lower than
the present background level, but it might become an important
contribution in future experiments. GEDEON (GErmanium
DEtectors in ONe cryostat) is a new project on WIMP detection using
larger masses of germanium of natural isotopic abundance~\cite{reviewmorales}. 
It will use the technology developed for the IGEX experiment and consist
of a set of $\sim$\,1\,kg germanium crystals (total mass of about
28$\,$kg), placed together in a compact structure inside one only
cryostat. This approach could benefit from the anticoincidence between
the crystals and a lower components/detector mass ratio to further
reduce the background with respect to IGEX. In this project 
a more efficient veto system should be designed to minimize
the contribution of the neutrons induced by muons in the shielding.

\section{Muon-induced neutrons in the surrounding rock}

Neutrons produced by  muons in the last few meters of the
underground laboratory rock are much less abundant that neutrons
coming from radioactivity. However, they could be relevant because
of their much higher energies. In addition, most of the
muon-induced neutrons in the rock are not detected by the veto
system and become as important as neutrons induced in the
shielding itself, even though these are produced much closer to
the detector. Following similar steps to those of the previous
section, we have first studied the neutron outcome after the muons
have traversed the overburden rock. We have performed a FLUKA
simulation with a muon spectrum corresponding to 2450\,m.w.e.\
traversing a layer of rock thick enough (10 meters) since neutrons
produced farther inside the rock would never reach laboratory
walls. The total neutron yield
is $4.6\times 10^{-4}$\,(g/cm$^{2}$)$^{-1}$ per muon, compatible
with values shown in other references \cite{wulandari,kudry,khal}.

\begin{figure}[htb]
\centerline{\includegraphics[angle=270,width=14cm]{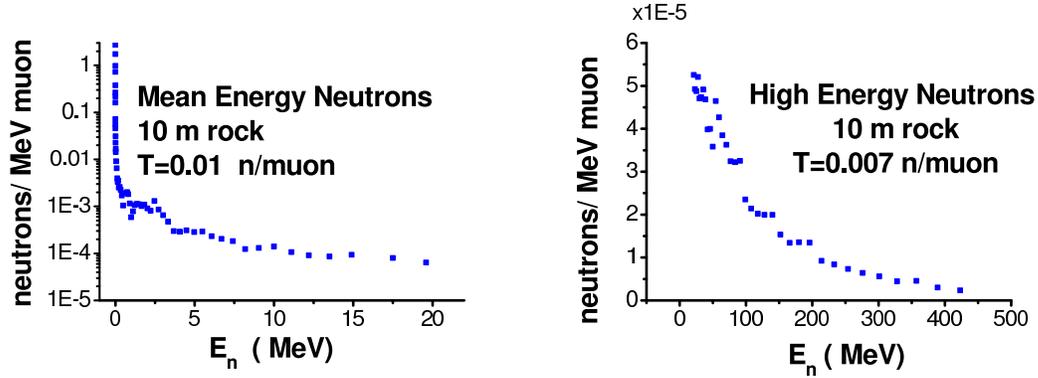}}
%\vspace{-1cm}
\caption{Energy spectrum of outgoing muon-induced neutrons in rock
according to the FLUKA simulation (squares).} \label{rockfluka}
\smallskip
\end{figure}

The relevant parameter we are interested in is, however, the
outgoing neutron outcome. In Fig.~\ref{rockfluka} we represent
both outgoing neutrons populations, HE and ME. We will no longer
be concerned about ME neutrons (0.01 neutrons per muon) since,
like those more abundant coming from radioactivity, they are
easily shielded as we will show in the next section. Therefore we
will just focus on HE ($>$\,20\,MeV) outgoing neutrons whose yield
results in a lower value of 0.007 neutrons per muon. This allows
to compute the total muon-induced neutron flux coming out of the
rock using the muon flux of the previous section. The estimated
result is $(1.73\pm 0.22 (stat) \pm 0.69 (syst))\times
10^{-9}$\,cm$^{-2}$\,s$^{-1}$.

\begin{figure}[htb]
\centerline{\includegraphics[width=10cm,angle=0]{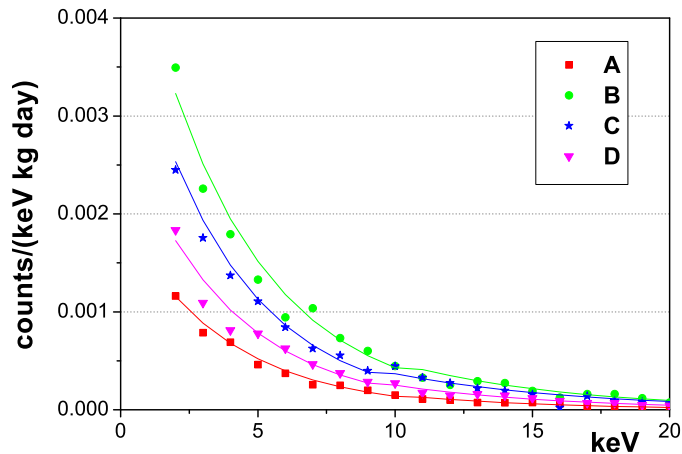}}
\vspace{-0.3cm} \caption{Energy spectra deposited in the detector
by muon-induced neutrons in rock at a depth of 2450 m.w.e.:
simulation for the different set-ups. Fits to exponential curves have
been drawn to guide the eye.} \label{nrock} \bigskip
\end{figure}

High-energy neutrons with the spectral shape of
Fig.~\ref{rockfluka} [which follows the analytical form of
Eq.~(\ref{muinhe})] have been launched from the outer shielding
surface for the different thicknesses of moderator of sets A--D to
simulate their transportation and interaction through the IGEX
geometry. The results of these simulations are shown in
Fig.~\ref{nrock} and the integrated values in Table \ref{ritmos2}.

\begin{table}[htb]
\centering \caption {Simulated background rates in IGEX-DM for the
energy region from 4 to 10\,keV due to $\mu$-induced neutrons in
the rock for the different set-ups for a neutron flux of $1.73
\times 10^{-9}$\,cm$^{-2}$\,s$^{-1}$. Only statistical errors are quoted.}
\label{ritmos2} \vskip 0.5cm
\begin{tabular}{|c|c|} \hline
 & counts/(keV\,kg\,day)\\ \hline

A &0.00041(1) \\ \hline

B &0.00124(4) \\ \hline

C &0.00091(4) \\\hline

D &0.00061(3) \\ \hline

%lead shield & $3.2\times 10^{-6}$ & 0.0011?? & 0.0025?? \\ \hline
\end{tabular}
\bigskip
\end{table}

It is interesting to see that the contribution is maximum for the
intermediate value of 20\,cm of neutron moderator, decreasing for
thicker walls or in the case of total absence of polyethylene:
very energetic neutrons (with some hundreds of MeV), which produce
nuclear recoils corresponding mostly to energies higher than our
range of interest in the absence of moderator walls, can be slowed
down in the presence of the polyethylene shielding, inducing in
this way nuclear recoils in the low-energy region. In any case,
the contribution of these neutrons to the IGEX-DM background is much
lower than the present background level.

\section{Neutrons from radioactivity in the surrounding rock}

\begin{figure}[tb]
\centerline{\includegraphics[width=7cm,angle=270]{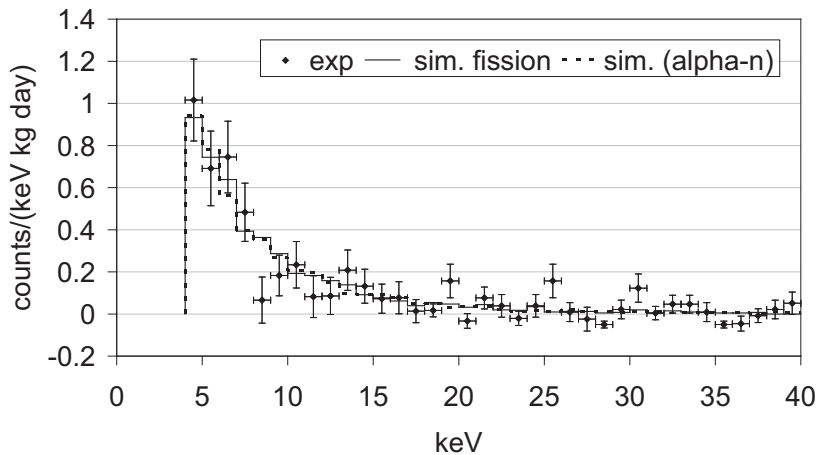}}
\caption{Comparison of the simulated spectra of
electron-equivalent energy deposited in the detector due to
fission and ($\alpha$,~n) processes in the rock assuming no neutron
moderator shielding (labelled \emph{sim. fission} and 
\emph{sim.\ (alpha-n)} respectively) 
with the difference between measured spectra for
sets A and C (labelled $exp$).} \label{fissionspectra} \smallskip
\end{figure}

Neutrons can be produced in the rock by spontaneous fission of
uranium and ($\alpha$,~n) reactions. The intensity of the
corresponding outgoing neutron flux is therefore dependent on the
kind of rock. In fact, several estimates of this flux in deep
underground locations give values in a range from $10^{-6}$ to
$10^{-5}$\,cm$^{-2}$\,s$^{-1}$
~\cite{chardin,perera,Heusser,wul2,belli,rindi,chazal,abdura,kim,hashemi}.
To sample the incident energies of fission neutrons a typical
fission spectrum has been used (energy $E$ is expressed in MeV):
\begin{equation}
\frac{dN}{dE} \propto E^{1/2} \exp(-E/1.29)
\end{equation}
\noindent while the spectral sampling for neutrons from ($\alpha$,
n) reactions is performed following the calculated spectrum shown
in Fig.~11 of Ref.~\cite{chazal}, deduced for the Modane Laboratory.
These spectra can be used to
sample the incident energies since the spectrum does not suffer
any significant deformation after the neutrons have traversed
several meters of rock.
Although the neutron spectrum coming from ($\alpha$,~n) processes
is expected to be dependent on the type of rock 
(see Ref.~\cite{wul2} for an spectrum obtained at the Gran 
Sasso Laboratory), the recoil spectrum at the detector will turn out
to be little sensitive to the exact form of the input neutron spectrum
(see Table~\ref{ritmos} and Fig.~\ref{fissionspectra}).

The transport of this neutron flux produced by fission or
($\alpha$,~n) reactions has been simulated in the IGEX geometries
corresponding to the experimental sets A, B and C (0, 20 and
40\,cm of neutron moderator in the outer part of the shielding).
The integrated rate seen by the Ge detector in the energy region
from 4 to 10\,keV for each case is listed in Table~\ref{ritmos}.

\begin{table}[htb]
\centering \caption {Relative importance of neutrons coming from
the radioactivity of the rock: calculated background rates for the
different set-ups [expressed in counts/(keV\,kg\,day)] in the
energy region from 4 to 10\,keV assuming, either that all of them
have a fission spectrum (second column), or that they come from
($\alpha$,~n) reactions (third column). A normalization of the
input neutron flux of $3.82\times 10^{-6}$\,cm$^{-2}$\,s$^{-1}$
has been assumed (see text).} \label{ritmos} \vskip 0.5 cm
\begin{tabular}{|c|c|c|c|} \hline & cm of moderator

& fission &  ($\alpha$,~n) reactions  \\ \hline

A & $0$ & $6\times 10^{-1}$& $6\times 10^{-1}$  \\ \hline

B & $20$ & $5\times 10^{-3}$ &  $1\times 10^{-2}$  \\ \hline

C & $40$ & $\sim 8\times 10^{-5}$ &  $\sim 1 \times 10^{-4}$
\\\hline

%lead shield & $3.2\times 10^{-6}$ & 0.0011?? & 0.0025?? \\ \hline

\end{tabular}
\bigskip
\end{table}

Monte Carlo simulations of the propagation of neutrons through
typical shieldings show that 90\% of neutrons of 1\,MeV (5\,MeV)
moderate down to $E<0.1$\,eV after 12\,cm (22\,cm) of
polyethylene, and, in a shielding of 40\,cm of polyethylene,
99.7\% of neutrons of 5\,MeV are moderated down to 0.1\,eV (and
practically all neutrons of 1\,MeV). For neutrons of higher
energies the fraction which is moderated down to $E<0.1$\,eV
after, say, 40\,cm of water is $\sim$\,92\% (for neutrons of
10\,MeV), $\sim$\,83\% (for neutrons of 25\,MeV) and $\sim$\,50\%
(for neutrons of 50\,MeV). Therefore one can assume that, given
the energies of neutrons coming from radioactivity, 40\,cm of
polyethylene (or water) are enough to moderate their whole population.
This is indeed what Table \ref{ritmos} shows: with 40\,cm of
moderator the neutron events are reduced by 3 or 4 orders of
magnitude.

This fact can be used to find out the normalization of the input
neutron flux. The simulation in case A (no moderator) can be
directly compared with the spectrum obtained by subtraction of the
experimental spectra C (40\,cm of moderator) from A, under the
assumption that the 40\,cm polyethylene stops only neutrons coming
from radioactivity. This is not strictly true, but it is a
reasonable assumption taking into account the different magnitude
of the flux of neutrons from radioactivity and the muon-induced neutron
flux, as seen in the previous sections.
In Fig.~\ref{fissionspectra} the difference between the experimentally
obtained spectra C and A is plotted together with the simulated
fission and ($\alpha$,~n) neutron spectra, all three curves showing a
remarkable agreement in their spectral shapes. The normalization
of the simulation is fixed by minimizing the difference between
simulated and experimental spectra, and it turns to be virtually
the same for neutrons coming from fission or ($\alpha$,~n). The
fast neutron flux from radioactivity in the Canfranc rock is then
deduced to be $(3.82\pm 0.44)\times 10^{-6}$\,cm$^{-2}$\,s$^{-1}$
which is within the expected range. Only statistical errors are
quoted; the difference in the estimated flux when considering 
that all the neutrons have either the fission spectrum or that
corresponding to ($\alpha$,~n) processes is less than 2\%. 
The error due to the
presence of muon-induced neutrons in the rock is estimated to be 0.04\%,
taking into consideration the simulated contribution of these neutrons to
the counting rate of the experiment. 
We observe in 
Table~\ref{ritmos} that for sets B and C, having 20 and 40\,cm
of neutron moderator, the contribution of ($\alpha$,~n) reaction
neutrons is slightly higher than that of fission neutrons assuming
the same flux for both.

As a conclusion for the understanding of the IGEX-DM background,
it seems clear that this neutron flux is shielded by 40\,cm of
neutron moderator at our present level of sensitivity. In fact
the neutron flux is almost eliminated with just 20\,cm of
moderator. This agrees with the experimental background shown in
Table~\ref{sets} for the set-up B where there was a lack of two
moderator walls owing to the presence of dewars.

\section{Conclusions}

A complete quantitative study of the neutron environment in the
LSC has been performed. The analysis is focused on the IGEX Dark
Matter experiment, whose low energy raw background (with no
nuclear recoil discrimination) is the lowest ever achieved. The
study has consisted in a set of simulations compared with several
sets of experimental data taken with the IGEX detector in
different conditions of neutron shielding. In decreasing order of
importance, the neutron populations studied, whose estimated
fluxes are summarized in Table \ref{estflux}, are: neutrons coming
from radioactivity in the laboratory rock, neutrons induced by
muons in the lead shielding, and neutrons induced by muons in the
laboratory rock. Neutrons produced by radioactivity in the lead
shielding have been excluded from the analysis since they are
negligible at the present level of background of IGEX-DM.

For neutrons produced by the interaction of muons in the lead of
the shielding, the results of the simulations have been compared
with the experimental spectrum obtained in coincidence with the
vetoes, which has allowed to set the rate of production of these
neutrons in lead at the LSC depth at $(4.8 \pm 0.6 (stat) \pm 1.9
(syst))\times 10^{-9}$\,cm$^{-3}$\,s$^{-1}$ leading to a muon flux
of $(2.47 \pm 0.31 (stat) \pm 0.98 (syst))\times
10^{-7}$\,cm$^{-2}$\,s$^{-1}$. This number is in the range of the
expected muon flux at that depth, and is in perfect agreement with
previous measurements at the Canfranc site giving a flux of
$2\times 10^{-7}$\,cm$^{-2}$\,s$^{-1}$. The effect of these
neutrons in the experimental data is small thanks to the muon veto
system used [otherwise, this kind of events would contribute to
0.16\,counts/(keV\,kg\,day) in the 4-10 keV region]. Improvements
in the efficiency of the muon veto system might be necessary to
reduce the effect of these neutrons in more sensitive experiments.

Regarding the neutrons coming from radioactivity in the rock, the
simulations have been compared with experimental data taken with
different thicknesses of neutron moderator, which has allowed us
to set the flux of this kind of neutrons to $(3.82 \pm 0.44)\times
10^{-6}$\,\cmds. We have concluded that the recent improvement
reported in our article \cite{igex2001} is compatible with the
complete rejection of this kind of neutrons that previously (in
\cite{igex2000}) accounted up to 50\% of the low energy data. With
40\,cm of neutron moderator, these neutrons should contribute to
the experimental data in less than $10^{-4}$
counts/(keV\,kg\,day).

Finally, with respect to the muon-induced neutrons in the rock,
they have been simulated consistently with the previous
information. From the incoming flux of muons, a flux of high
energy neutrons going out of the rock of $(1.73 \pm 0.22(stat) \pm
0.69(syst))\times 10^{-9}$\,cm$^{-2}$\,s$^{-1}$ is obtained, which
finally contribute up to $10^{-4}$ counts/(keV\,kg\,day) to the
IGEX-DM background, a number much below the present experimental
level.

Briefly, the neutrons can be rejected as responsible for the low
energy events that populate the last IGEX-DM data below
$\sim$\,20\,keV substantially over the expectations (see
\cite{igex2001}) and whose identification was partially the
motivation of the present work. The information gathered by the
present analysis will be also extremely useful in the design of
future experiments in the LSC.

\begin{table}[htb]
\centering \caption {Results for the estimates of the fluxes of
different neutron populations reaching the IGEX-DM experimental
setup in the LSC.} \label{estflux} \vskip 0.5 cm {\scriptsize
\begin{tabular}{|c|c|} \hline

neutrons from radioactivity of the rock & $(3.82 \pm 0.44)\times
10^{-6}$\,cm$^{-2}$\,s$^{-1}$\\ \hline

muon-induced neutrons in the rock  &  $(1.73 \pm 0.22(stat) \pm
0.69(syst))\times 10^{-9}$\,cm$^{-2}$\,s$^{-1}$ \\ \hline

muon-induced neutrons in the shielding lead & $(4.8 \pm 0.6 (stat)
\pm 1.9 (syst))\times 10^{-9}$\,cm$^{-3}$\,s$^{-1}$ \\\hline

%lead shield & $3.2\times 10^{-6}$ & 0.0011?? & 0.0025?? \\ \hline

\end{tabular}
} \bigskip
\end{table}

\section{Acknowledgments}

During the writing of this article, Prof. Angel Morales passed
away. Even if it is impossible to express here how all of us are
indebted to him, we wish to show our deepest and grateful
acknowledgement not only for encouraging this study but also for
sharing with us his knowledge and enthusiasm for work.

The Canfranc Underground La\-bo\-ra\-to\-ry is operated by the
University of Zaragoza. This research was partially funded by the
Spanish Ministry of Science and Technology (MCYT) under contract
No. FPA2001-2437. We are indebted to our IGEX colleagues for their 
collaboration in the IGEX and IGEX-DM experiments.


\begin{thebibliography}{00}

\bibitem{map}
D.N.~Spergel {\it et al.}, Astrop.\ Phys.\ J.\ Suppl.\ 148 (2003)
175. See recent results of MAP on http://map.gsfc.nasa.gov/.
\bibitem{sn}
R.A.~Knop {\it et al.}, Astrop.\ Phys.\ J.\ 598 (2003) 102. See
recent results of SN project on http://panisse.lbl.gov/.
\bibitem{sdss}
M.~Tegmark {\it et al.}, astro-ph/0310723 (accepted in Phys.\
Rev.\ D). See recent results of SDSS on http://www.sdss.org/.

\bibitem{reviewmorales} A.~Morales, Review Talk at the XXXth
International Winter Meeting on Fundamental Physics, Jaca, Huesca
(Spain), 2002, Nucl.\ Phys.\ B (Proc.\ Suppl.) 114 (2003) 39;
A.~Morales, Review Talk at the TAUP2003, Seattle (USA), to appear
in the Proceedings.

\bibitem{Silva}
A.~da Silva {\it et al.}, NIM A 354 (1995) 553.


\bibitem{igex2000}
A.~Morales {\it et al.} [IGEX Collaboration],
%``New constraints on WIMPs from the Canfranc IGEX dark matter search,''
Phys.\ Lett.\ B 489 (2000) 268. %[arXiv:hep-ex/0002053].

\bibitem{igex2001}
A.~Morales {\it et al.} [IGEX Collaboration], Phys.\ Lett.\ B 532
(2002) 8.
%``Improved constraints on WIMPs from the international germanium  experiment IGEX,''
%[arXiv:hep-ex/0110061].

\bibitem{wavelets}
I.G.~Irastorza {\it et al.}, Astrop.\ Phys.\ 20 (2003) 247.

\bibitem{cdms}
R.~Abusaidi {\it et al.} [CDMS Collaboration], Phys.\ Rev.\ Lett.\
84 (2000) 5699; D.~Abrams {\it et al.} [CDMS Collaboration],
Phys.\ Rev.\ D 66 (2002) 122003; D.~Akerib {\it et al.} [CDMS
Collaboration], Phys.\ Rev.\ D\ 68 (2003) 082002.

\bibitem{edelweiss}
A.~Benoit {\it et al.} [EDELWEISS Collaboration], Phys.\ Lett.\ B
513 (2001) 15; A.~Benoit {\it et al.} [EDELWEISS Collaboration],
Phys.\ Lett.\ B 545 (2002) 43.

\bibitem{cresst}
M.~Bravin {\it et al.} [CRESST Collaboration], NIM A 444 (2000)
323; G.~Angloher {\it et al.} [CRESST Collaboration], Astrop.\
Phys.\ 18 (2002) 43.

\bibitem{rosebud}
S.~Cebri\'{a}n {\it et al.}, Phys.\ Lett.\ B 563 (2003) 48.

\bibitem{chardin}
G.~Chardin, Proceedings of the Fourth International Workshop on
Identification of Dark Matter 2002, York (England), World
Scientific p.~470.

\bibitem{wulandari}
H.~Wulandari, Proceedings of the Fourth International Workshop on
Identification of Dark Matter 2002, York (England), World
Scientific p. 464; H. Wulandari {\it et al}, hep-ex/0401032.

\bibitem{kudry}
V.A.~Kudryavtsev, N.J.C.~Spooner, and J.E.~McMillan, NIM A 505
(2003) 683; V. A. Kudryavtsev, Proceedings of the Fourth
International Workshop on Identification of Dark Matter 2002, York
(England), World Scientific p. 477.

\bibitem{perera}
T.A.~Perera, PhD thesis, Case Western Reserve University (2002).\\
http://cosmology.berkeley.edu/preprints/cdms/Dissertations/tap\_thesis.pdf

\bibitem{igexdbd}
C.E.~Aalseth {\it et al.}, Phys.\ Rev.\ C 59 (1999) 2108;
D.~Gonz\'{a}lez {\it et al.}, Nucl. Phys. B (Proc.\ Suppl.) 87
(2000) 278; C.E.~Aalseth {\it et al.}, Phys.\ Rev.\ D 65 (2002)
092007.

\bibitem{Heusser} G. Heusser, Ann.\ Rev.\ Nucl.\ Part.\ Sci.\ 45 (1995)
543.

\bibitem{geant4}
S.~Agostinelli {\it et al.} [GEANT4 Collaboration], NIM A 506
(2003) 250.

\bibitem{fluka}
A.~Fass\`{o}, A.~Ferrari, P.R.~Sala, ``Electron-photon transport
in FLUKA: status", Proceedings of the MonteCarlo 2000 Conference,
Lisbon, October 23--26 2000, A.~Kling, F.~Barao, M.~Nakagawa,
L.~Tavora, P.~Vaz eds., Springer-Verlag Berlin, p.~159-164 (2001);
A.~Fass\`{o}, A.~Ferrari, J.~Ranft, P.R.~Sala, ``FLUKA: Status and
Prospective for Hadronic Applications", ibid, p.~955-960 (2001).

\bibitem{golwala}
S.R.~Golwala, PhD thesis, University of California at Berkeley (2000).\\
http://cosmology.berkeley.edu/preprints/cdms/golwalathesis/

\bibitem{Lipary}
P.~Lipari {\it et al.}, Phys.\ Rev.\ D 44 (1991) 3543.

\bibitem{khal}
F.F.~Khalchukov {\it et al.}, Il Nuovo Cimento 6C (1983) 320;
F.F.~Khalchukov {\it et al.}, Il Nuovo Cimento 18C (1995) 517.

\bibitem{wul2}
H. Wulandari {\it et al.}, hep-ex/0312050.

\bibitem{belli}
P.~Belli {\it et al.}, Il Nuovo Cimento A 101 (1989) 959.

\bibitem{rindi}
A.~Rindi {\it et al.}, NIM A 272 (1988) 871.

\bibitem{chazal}
V.~Chazal {\it et al.}, Astrop.\ Phys.\ 9 (1998) 163.

\bibitem{abdura}
J.N.~Abdurashitov {\it et al.}, Nucl. Phys. B (Proc. Suppl.) 110
(2002) 320.

\bibitem{kim}
H.J.~Kim {\it et al.}, Astrop.\ Phys. 20 (2004) 549.

\bibitem{hashemi}
S.R.~Hashemi-Nezhad, L.S.~Peak, NIM A 357 (1995) 524.
\end{thebibliography}
\end{document}